# Magnetic Tunnel Junction Random Number Generators Applied to Dynamically Tuned Probability Trees Driven by Spin Orbit Torque

Andrew Maicke[1], Jared Arzate[1], Samuel Liu[1], Jaesuk Kwon[1], J. Darby Smith[2], James B. Aimone[2], Shashank Misra[2], Catherine Schuman[3], Suma G. Cardwell[2], and Jean Anne C. Incorvia[1]

[1] University of Texas at Austin, Austin, TX, USA
[2] Sandia National Laboratories, Albuquerque, NM, USA
[3] University of Tennessee, Knoxville, Knoxville, TN, USA

E-mail: incorvia@utexas.edu

**Abstract**

Perpendicular magnetic tunnel junction (pMTJ)-based true-random number generators (RNG) can consume orders of magnitude less energy per bit than CMOS pseudo-RNG. Here, we numerically investigate with a macrospin Landau-Lifshitz-Gilbert equation solver the use of pMTJs driven by spin-orbit torque to directly sample numbers from arbitrary probability distributions with the help of a tunable probability tree. The tree operates by dynamically biasing sequences of pMTJ relaxation events, called 'coinflips', via an additional applied spin-transfer-torque current. Specifically, using a single, ideal pMTJ device we successfully draw integer samples on the interval [0,255] from an exponential distribution based on p-value distribution analysis. In order to investigate device-to-device variations, the thermal stability of the pMTJs are varied based on manufactured device data. It is found that while repeatedly using a varied device inhibits ability to recover the probability distribution, the device variations average out when considering the entire set of devices as a 'bucket' to agnostically draw random numbers from. Further, it is noted that the device variations most significantly impact the highest level of the probability tree, with diminishing errors at lower levels. The devices are then used to draw both uniformly and exponentially distributed numbers for the Monte Carlo computation of a problem from particle transport, showing excellent data fit with the analytical solution. Finally, the devices are benchmarked against CMOS and memristor RNG, showing faster bit generation and significantly lower energy use.

## 1. Introduction

The generation of large, high-quality sets of random numbers is critical for many modern computing purposes, such as the modeling of quantum or high-energy physics systems, precise simulation of climate models, optimization, cryptography, and others. Traditionally, this has been achieved using deterministic computing architecture, which includes von Neumann machines such as clusters of graphics processing units or central processing units. This has two implications for the production of random numbers. First, the numbers are not truly stochastic in nature, and thus are more accurately referred to as 'pseudo'-random numbers. And second, significant energy and computational resources are required to appropriately emulate stochastic behavior using only deterministic inputs and operations on CMOS technology [1,2].

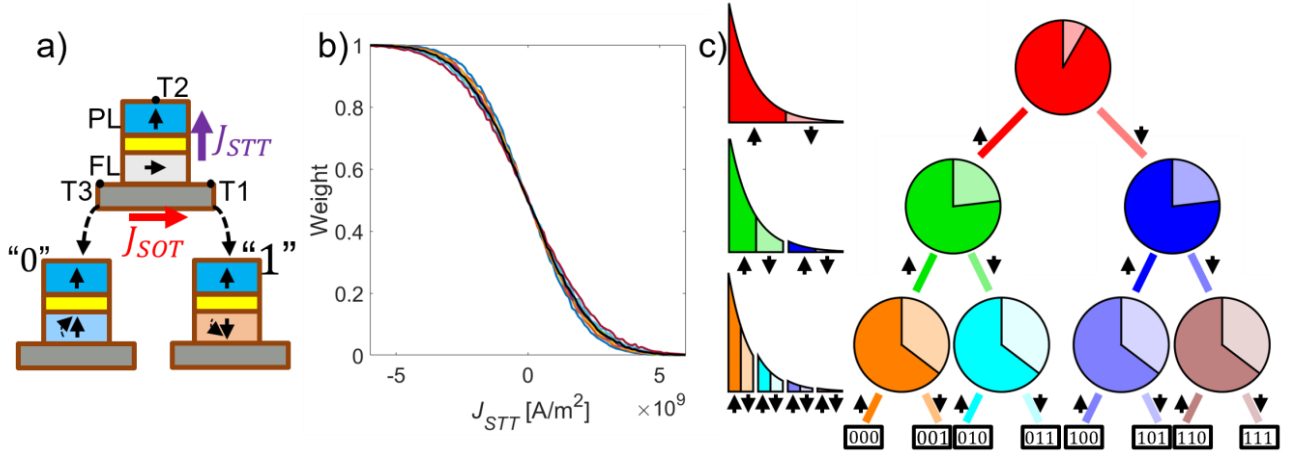

Fig. 1. a) PMA MTJ stack showing FL being held in-plane via applied SOT current between T1 and T3, and relaxing to P or AP states upon SOT current removal. Relaxation can be biased via application of STT current between T1 and T2. b) Calculated ideal MTJ device S-curve shown in black, mapping desired weighting to applied STT current. Colored curves show S-curve distortion as MTJ parameters vary. c) Mechanics of the tunable probability-tree, showing a three-level tree operation with associated PDF and results as an example. Because the example distribution is an exponential, the weightings across branches on a given level are equal; for arbitrary PDFs this will not be the case.

Probabilistic approaches to computing have gained interest in recent years as potential solutions to these problems [3,4]. For example, it is possible to leverage the inherent stochasticity of nanoscale devices such as the magnetic tunnel junction (MTJ) to produce probabilistic behavior at the device level, without needing complex mathematical functions or computer hardware to convert deterministic inputs into pseudo-random outputs [5-8]. This allows for 'true-random number' generation (tRNG), where the resulting random bitstreams are the results of directly sampling stochastic thermal processes.

The MTJ has been studied as a stochastic bit using various knobs to sample or control its stochasticity. An individual sampling of an MTJ is treated as a 'coinflip', i.e. the MTJ state at a given time is interpreted to be in one of two binary positions 'heads' and 'tails'. The MTJ p-bit has a low energy barrier between its two resistance states, sampled as it thermally fluctuates in time; spin transfer torque switching of the MTJ free layer is stochastic, and choosing the correct switching current can produce a 50:50 probability of switching. To avoid dependence on temperature, instead the MTJ energy barrier can be lowered and then raised in time to produce random binary bits, evaluated in our previous work [9]. The probability for the MTJ to be in either state when sampling can be controlled via direct application of spin transfer torque, spin orbit torque, voltage-controlled magnetic anisotropy [5,6], or a combination thereof; by taking a sequence of coinflips with appropriate biasing between samples, it is possible to assemble sets of true-random numbers.

While using MTJs to produce bitstreams of true-random numbers has been investigated, a less explored property of the MTJ as an RNG is the ability to dynamically tune the MTJ switching probability. This opens up the ability to sample from arbitrary probability distributions with the help of a probability-tree [10]. This framework allows direct sampling of arbitrary distributions without needing to spend resources converting uniformly-distributed random data or use techniques such as rejection sampling to ensure bit quality, making it more energy efficient.

In this work, we use physics-based simulations to demonstrate the use of dynamically tunable MTJ RNGs for sampling from a probability distribution. A probability tree for the exponential distribution is chosen due to the usefulness ubiquity of such a distribution in physics applications [11-13]. Using a single, ideal device it is shown via p-value analysis that the drawn samples are reasonably sampled from the desired distribution. Since the precision of such a probability distribution is essential to many of the applications, thermal energy cycle-to-cycle variation as well as device-to-device variation of state-of-the-art fabricated MTJ arrays [14,15] are studied and the resulting effects on true-random number generation are shown. Specifically, when using a single, varied device the ability to accurately draw samples from the desired distribution is diminished when the variation gets too large; fortunately, this can be rectified by using a collection of varied devices to generate the entire random dataset. Finally, the framework is used to draw exponential and uniform random numbers for use in a Monte Carlo physics simulation, where excellent fit with the analytical data is shown. Comparing to CMOS pRNG and tRNG, several orders of energy savings are achieved.

The paper is divided as follows: Section 2 describes the physical operation of the MTJ coinflip device in the probability tree to construct a random number pulled from a prescribed probability distribution. Section 3 describes the numerical model used for this investigation and the particular material and device properties simulated. Section 4 uses the ideal MTJ-based device to generate 8-bit integer samples from an exponential distribution. Section 5 introduces realistic device-to-device variations and analyzes the effects on the resulting sampled distributions, as well as mitigation strategies. Section 6 applies the described framework to random walk solutions for the computation of particle absorption. And finally, Section 7 concludes the paper.

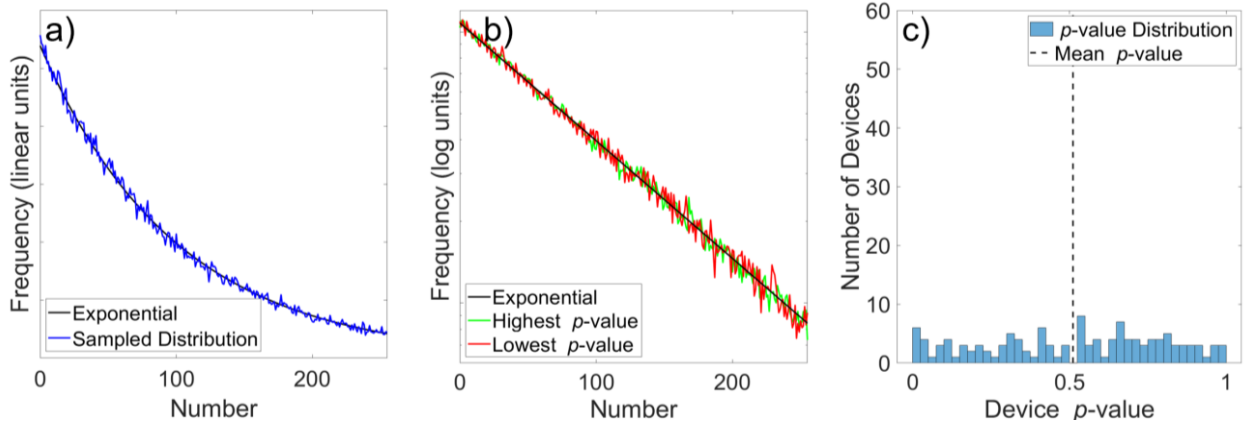

Fig. 2. a) Sampled frequencies, in linear units, of sampled numbers using the MTJ with a probability-tree compared to the exponential distribution. b) Repeated experiments on a log unit scale with the highest and lowest $p$-value datasets. c) Distribution of $p$-values across all 100 experiments.

## 2. MTJ Operation in the Probability Tree

A single MTJ is a nanoscale device with three critical layers: a ferromagnetic free layer (FL), an insulating layer, and a ferromagnetic pinned layer (PL) as shown in Fig. 1a with perpendicular magnetic anisotropy (PMA). The magnetization of the PL is held fixed, while the FL is allowed to rotate. When the magnetizations of the FL and PL are parallel (P), ie. they point in the same direction, the MTJ is in a low resistance state. When the magnetizations of the two layers are anti-parallel (AP) the resistance of the MTJ is in a high resistance state. With an energy barrier between the states greater than thermal energy, each state is a local energy minimum, and the MTJ will remain in its state until application of an external signal.

Considering the P and AP states as the two possible binary states, it is possible to construct an unbiased coinflip device by pulsing a spin-orbit current in a heavy metal layer beneath the FL between T1 and T3 as shown in Fig. 1a. This rotates the magnetization of the FL to be in-plane, and perpendicular to that of the PL via the spin Hall effect [16]. Removing the spin-orbit current will cause the FL to stochastically relax back into either the P or AP states with equal probability at finite temperatures. It is also possible to create a biased coinflip by applying a spin-polarized current between T1 and T2; the precise amplitude of the spin-polarized current sets the probability to fall into either the P or AP states. The relative weight given to each state as a result of the applied spin-polarized current is described by the sigmoidal device 'S-curve' [17], calculated in Fig. 1b.

TABLE I
SIMULATION PARAMETERS

| Symbol | MTJ Parameter | Value |
|---|---|---|
| $\alpha$ | Gilbert damping | 0.03 |
| $M_S$ | Saturation magnetization | $6 \times 10^5 \, A/m$ |
| $\Delta$ | Thermal stability factor | 70 |
| $A_{ex}$ | Exchange stiffness | $4 \times 10^{-6} \, erg/cm$ |
| $RA$ | Resistance-area product | $7 \, \Omega \cdot \mu m^2$ |
| $TMR$ | Tunneling magnetoresistance ratio | 150% |
| $a$ | MTJ diameter | $50 \times 10^{-9} \, m$ |
| $t_{FL}$ | Free layer thickness | $1.1 \times 10^{-9} \, m$ |
| $t_{MgO}$ | MgO layer thickness | $1.5 \times 10^{-9} \, m$ |
| Symbol | Device Operation Parameter | Value |
| $T_{pulse}$ | Pulse time | $10 \times 10^{-9} s$ |
| $T_{relax}$ | Relax time | $15 \times 10^{-9} \, s$ |
| $J_{SOT}$ | SOT current density | $5 \times 10^{11} A/m^2$ |
| $P$ | Spin polarization of tunnel current | 0.6 |
| $T$ | Temperature | $300 \, K$ |

Interpreting the outputs of a series of unbiased coinflips specifically as the digits of a binary number, i.e. as 0's and 1's, allows for generation of true-random integers sampled from a uniform distribution. In order to sample from more complex distributions, a probability-tree can be used to properly tune successive coinflips as shown in Fig. 1c. Selecting reasonable lower and upper bounds to draw random samples from, the bias of the first coinflip is determined by dividing the probability distribution function (PDF) of the target distribution at the midpoint of the chosen bounds and the area-fractions underneath the PDF curve to the left and right of the division (readily calculable using the cumulative distribution function) represents the probability for the MTJ to relax into the P or AP states, respectively. This desired weighting is mappable to an applied spin-torque transfer (STT) current via the device ideal S-curve. For the following coinflip, the PDF section corresponding to the result of the first flip (left side for P, right side for AP) is again split into two parts and used to determine the next bias. This process continues recursively down the tree until all desired bits have been sampled. Each leaf thus corresponds to a bin centered on the final drawn value; the width and number of available bins depends on the choice of initial lower and upper bounds, and on the number of levels in the probability-tree. Fig. 1c illustrates the process for generating random integers on the interval [0, 7] from an exponential distribution.

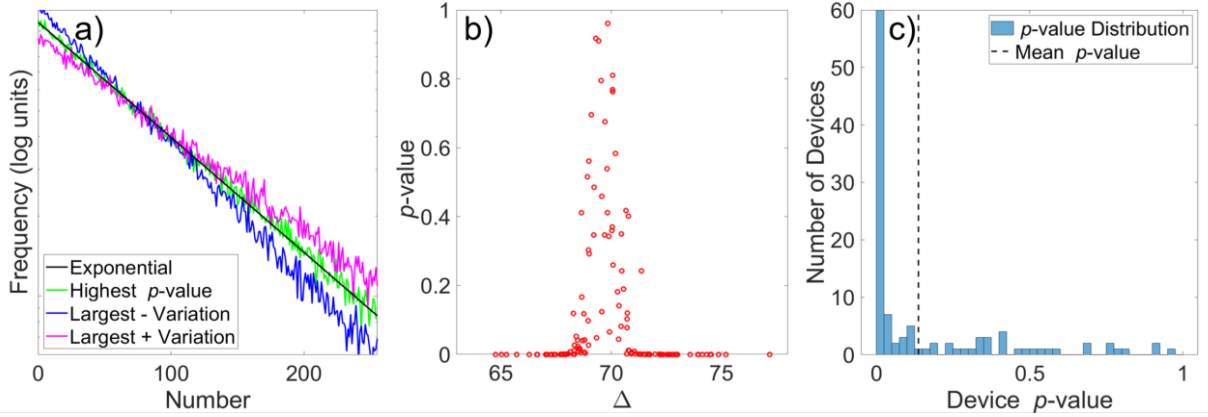

Fig. 3. a) Sampled frequencies, in logarithmic units, for the experiments with the largest positive and negative Δ deviation. b) *p*-values for all device deviation experiments, showing workable deviation limits. c) *p*-value distributions for the varied device experiments.

While a particular MTJ device can be characterized by an ideal S-curve that allows for accurate biasing with a STT current, inevitably device variations will be introduced in the manufacturing process [14]. Geometric imperfections, uneven growth rates, and other factors can cause device quantities such as the TMR, resistance-area product (RA), and thermal stability factor (Δ) to vary between devices fabricated on the same die. This leads to variations in the device S-curves, calculated as the colored lines in Fig. 1b by varying Δ within the range ±3%. As a result of these parameter variations, the sampled random bitstreams from these non-ideal devices will become biased [18, 19], and the random variables sampled via the probability-tree may not belong to the intended target distribution. The impact of parameter deviations on the devices, as well as the result of using collections of devices with normally distributed variations to generate sets of randomly generator numbers, is studied in section 5.

## 3. Numerical Model

To investigate the ability for SOT-driven MTJ devices to accurately sample numbers from arbitrary probability distributions using a tuned probability-tree, the FL magnetization is modeled as a 3-dimensional vector $\vec{m}$ obeying the time-dynamic Landau-Lifshitz-Gilbert equation (LLG) under a macrospin approximation. Including both STT and SOT terms, the LLG is written as:

$$\frac{\partial \vec{m}}{\partial t} = -\gamma \vec{m} \times \vec{H}_{eff} + \alpha \vec{m} \times \frac{\partial \vec{m}}{\partial t} - \beta P J_{STT} |\gamma| \vec{m} \times (\vec{m} \times \vec{m}_{PL}) - \beta \eta J_{SOT} \vec{m} \times (\vec{m} \times \vec{\sigma}_{SOT}) \quad (1)$$

where $P$ is the spin-polarization, $\gamma$ the gyromagnetic ratio, $\alpha$ the Gilbert damping constant, $\vec{m}_{PL}$ the magnetization of the PL, $J_{STT}$ the amplitude of the biasing STT current density, and $J_{SOT}$ the amplitude of the SOT current density through the heavy-metal layer, and $\vec{\sigma}_{SOT}$ its direction. $\beta$ is defined as $\beta = \sqrt{\gamma \hbar / 2 q_e t_f M_S}$ where $q_e$ is the electron charge, $t_f$ the FL thickness, and $M_S$ the saturation magnetization of the FL and PL. The term $\vec{H}_{eff}$ is an effective field term that includes both $\hat{z}$ directed PMA and stochastic thermal effects. Equation (1) is discretized in time and solved with a time-integration scheme. The device simulated is a typical CoFeB PMA MTJ with MgO tunnel barrier. Device parameters such as RA, TMR, and Δ are based on data from fabricated devices in [14]; note that in order to maintain PMA under these conditions, the saturation magnetization (unreported in [14]) had to be sightly lowered past typical values, but is kept within realistic physical limits [20]. The full set of device and simulation parameters is listed in Table 1, however other PMA devices are expected to work with the tuned probability-tree framework provided accurate characterization of the ideal S-curve.

For demonstration of the method, integer samples are drawn on the interval [0, 255]. This requires 8 coinflips per sample, with each coinflip interpreted as the bit of an 8-bit integer number. The chosen probability distribution to draw random samples $x$ from is the exponential distribution, whose PDF is:

$$f(x; \lambda) = \lambda e^{-\lambda x}, x \geq 0 \quad (2)$$

where $\lambda > 0$ is the rate parameter. For this work, $\lambda$ is chosen to be 0.01, resulting in 92.19% of the expected draws to lay in the interval [0, 255]. As the exponential distribution is unbounded on the positive side of the real number line, there is some expected distribution error as a result of truncation of the sample domain, which will be explored in section 6.

To capture MTJ device variations, the thermal stability factor Δ is randomly varied about the value 70 by ±3% following characterization of fabricated devices in [14]. The

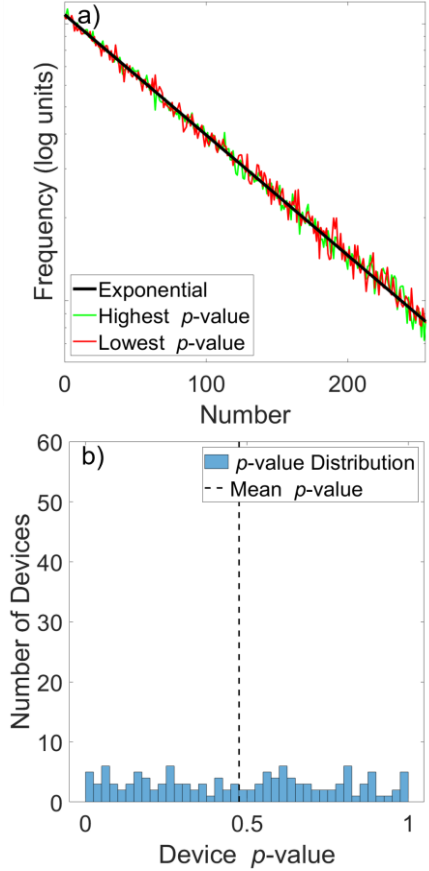

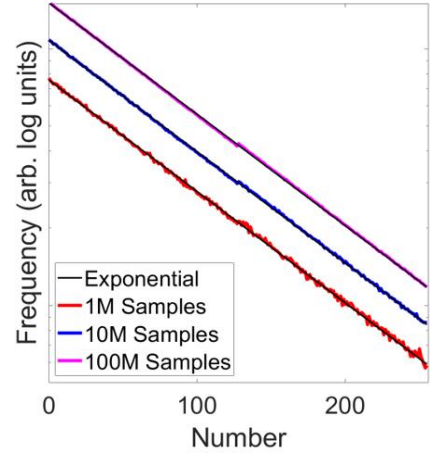

Fig. 5. Sampled distributions, in logarithmic units, when generating datasets of sizes 1M, 10M and 100M using the device-variation agnostic approach. Note the three cases are arbitrarily offset from each other in the y-axis to show in a single plot.

$$\chi^2 = S \sum_{i=1}^{N} \frac{(O_i - E_i)^2}{E_i} \quad (4)$$

where $N = 256$ is the number of possible output numbers (degrees of freedom), $O_i$ is the sampled (observed) frequency of each number, $E_i$ the expected frequency calculated from the exponential PDF, and $S$ the total number of samples drawn. The $\chi^2$ parameter is therefore a sum of squared deviations between the observed and expected results; a small $\chi^2$ implies little deviation, while a large $\chi^2$ implies there is a meaningful difference between the observed data $O_i$ and the expected data $E_i$. The $\chi^2$ parameter is used with (one-sided, right-tailed) $p$-value testing to determine the probability of obtaining a result with at least as much deviation as that observed; ie, it is the expected likelihood that the result of an experiment would be at least as extreme as the observed data. $p$-values greater than 0.05 are generally considered to imply a good data-fit [23]. For the blue curve shown in Fig. 2a, $\chi^2 = 214.57$, corresponding to a $p$-value of 0.972 and thus can be considered to have successfully drawn observed samples from an exponential distribution.

However, it should be noted that the $p$-value itself is a random variable; repeated experiments will draw different observed data and therefore will compute different $\chi^2$ parameters corresponding to different $p$-values. Thus, the previous experiment is repeated 100 times; the distributions resulting from the experiments corresponding to the highest (best) and lowest (worst) $p$-values are compared to the exponential distribution on a log-scale in Fig. 2b. It is seen that even the 'worst' set of observed data is not easily visually discernable from an exponential distribution, or even the 'best' set of observed data, demonstrating how a single $p$-value is not enough to characterize the tuned probability-tree performance. Ideally, the distribution of $p$-values from a large set of repeated experiments should be uniform with mean 0.5 thermal stability factor is related to both the energy barrier $E_b$ and the anisotropy energy density $K_u$ through the relation [21],

$$\Delta k_B T = E_b = K_u V \quad (3)$$

where $k_B$ is the Boltzmann constant, $T$ the device temperature and $V$ the FL volume. Certain other device variations, such as the MTJ critical dimension (MTJ diameter) are also expected to affect device performance, however we find $\Delta$ has the most significant effect on the RNG behavior; see [22] for such an exploration of other parameters using reinforced learning.

Fig. 4. a) Highest and lowest $p$-value datasets taking a device agnostic approach when drawing random samples. b) $p$-value distribution for the device agnostic experiments.

## 4. Drawing Samples

Fig. 2 shows the result of using the ideal MTJ device to draw 100k integer samples from an exponential distribution using the probability-tree. The blue line in Fig. 2a represents the relative frequency of the sampled draws for each integer in the set [0,255]. It can be seen that, with some apparent noise due to the random nature of the sampling, the observed MTJ-sampled distribution frequency closely follows that of the expected distribution. In order to evaluate the quality of the sampled distribution, a $\chi^2$ parameter is computed via [24,25]. The spread of $p$-values for the 100 performed experiments is captured in the histogram in Fig. 2c, which shows that the $p$-value distribution has this desired behavior.

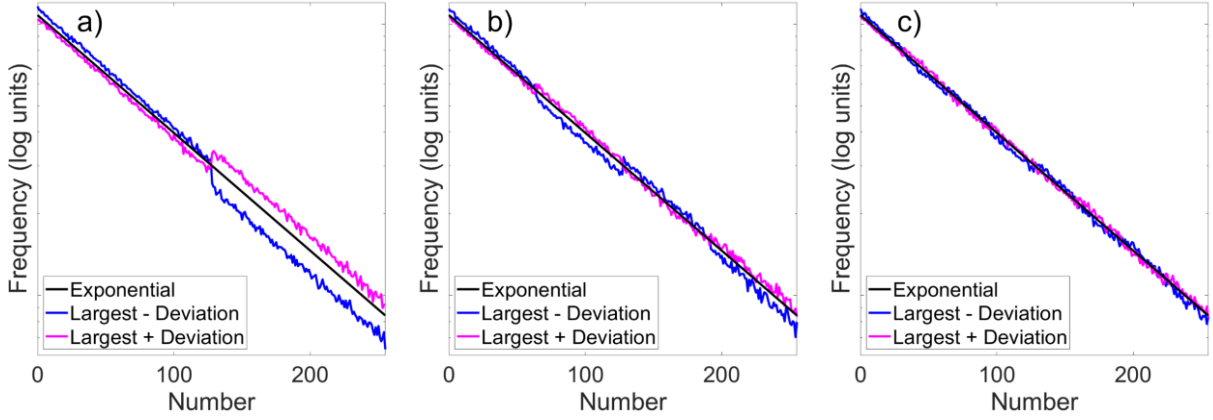

Fig. 6. a) Largest positive and negative deviation dataset in the first coinflip (top of the probability tree) when not allowing deviations in the other coinflips. b) Largest positive and negative deviation in the second coinflip, keeping other levels unvaried. c) Largest positive and negative deviation in the third coinflip, keeping other levels unvaried.

## 5. Device-to-Device Variations

In practice, realizing the performance from the previous section would require either accurate knowledge of each individual device's S-curve or a set of perfect devices. However, as seen in Fig. 1b, when we include parameter deviations, the calculated S-curve becomes distorted from the ideal case. To investigate the effect of having device variations on the ability to sample from exponential distributions, Figures 3a-c show the results of 100 experiments where a single MTJ is used whose thermal stability is varied from the ideal according to the discussion in Section 2. Unlike the previous non-varying case, there is significant deviation in the quality of the observed data as the device parameters stray from the designed value. Specifically, in Fig. 3a, it can be seen that the device with the largest positive deviation under-samples low integers and over-samples high ones. The opposite behavior is observed for the device with the largest negative deviation. Each extreme is essentially sampling exponentially distributed numbers from distributions with a modified $\lambda$ parameter. The $p$-values for each experiment as a function of thermal stability are shown in Fig. 3b, where it can be seen that while the $p$-values are roughly uniformly distributed in the range $68 \leq \Delta \leq 71$, they are nearly always near 0 otherwise. The distribution of all $p$-values is shown in Fig. 3c, where it is obvious that they do not follow a uniform distribution nor do they have a mean of 0.5.

This implies that there is a certain variation tolerance which can be handled by the device under an assumed ideal S-curve; if the device parameters stray too far from the designed values, the applied STT currents will not accurately bias the individual flips. In this case, it is more accurately the *ratio* of thermal stability (anisotropy energy density) to saturation magnetization $K_u/M_s$ that determines the quality of the device. This ratio controls the level of PMA in the MTJ device, which is necessary for proper operation. To see this, consider (as a simplification) the total effective anisotropy only due to $K_u$ and saturation magnetization:

$$K_{eff} = K_u - \frac{\mu_0 M_s^2}{2}. \qquad (5)$$

If $K_{eff} > 0$, the device will exhibit PMA; if $K_{eff} < 0$, it will be IMA. Thus, to remain PMA it must be true that $K_u > \mu_0 M_s^2/2$. When the ratio becomes too small, the device more readily exhibits in-plane magnetic anisotropy (IMA), meaning the magnetization of the FL prefers to remain in-plane even without application of a SOT current and does not properly relax to either the P or AP states. Instead, the observed data increasingly becomes the result of the (uniformly-distributed) thermal noise. When the ratio is too large, the SOT current cannot fully bring the FL magnetization to the perpendicular state even with assistance from the biasing STT current, and thus is likely to remain in whatever state it already happens to be in. Thus when using a single device to generate a large set of random numbers, it is critical to get the device parameters correct.

Instead of repeatedly using a single (potentially non-ideal) device to draw an entire set of random numbers in this manner, it may be instead preferable to draw each new number from a random device. In this mode of operation, the user is 'agnostic' as to where the number comes from; they do not know and do not care ahead of time which MTJ device will generate the number. This operation strategy may be preferable as while an MTJ can theoretically have unlimited endurance [26,27], in practice MTJ devices are expected to have a cycle life of $\sim 10^{15}$ cycles before degradation [28], and thus distributing the work can let devices last longer. Further, each device can generate numbers in parallel, instead of needing to sequentially be sampled millions or billions of times.

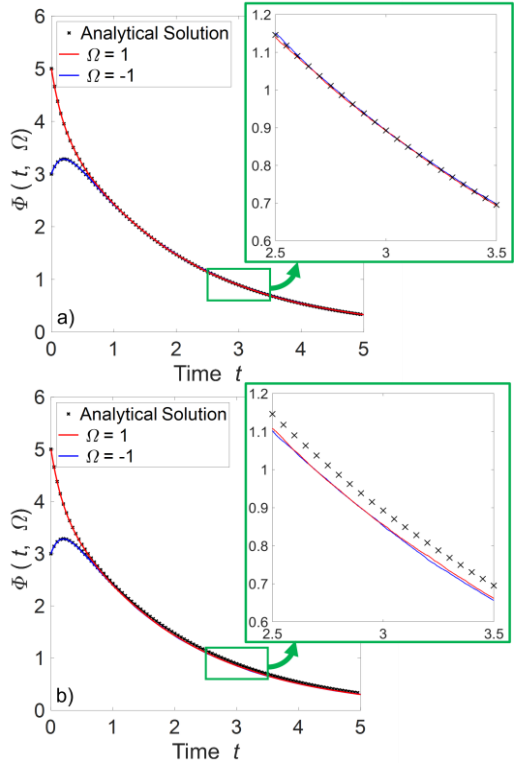

Fig. 7. a) Averaged result of 128k random walk simulations using the MTJ-based device and tuned probability-tree for simulating particle transport. b) Truncation of the sampled distribution domain to the interval [0,1] causes underestimation of the true solution, due to the absence of rare, large draws.

Figures 4a-b show another benefit of this approach. Here, the varied devices are again used to generate sets of 100k random numbers, but this time each sampled number comes from a unique (randomly-sampled) device with a random deviation in thermal stability. In Fig. 4b, it is seen that the $p$-value distribution again becomes uniform with a mean centered at 0.5. This is because, assuming the variations in the devices are normally-distributed and centered around the device parameters chosen to correspond to the ideal S-curve, there is an equal likelihood for any device to be under- or over-biased compared to the desired bias level. The net result is that the collection of devices as a whole *does* reasonably sample from an exponential distribution. As before, it is seen in Fig. 4a that the differences between the observed best, observed worst, and exactly exponential distribution results are visually indiscernible.

Using this strategy, it is possible to observe the shapes of the distributions as the number of pulled random samples increases. Fig. 5 shows the distribution as the number of samples is increased to 1M, 10M, and 100M. At these higher sample sizes, the visible random fluctuations present in the set of 100k samples disappears. Despite this, the $\chi^2$ parameters computed via (4) for each case start to diverge as the sample size $S$ increases. This is a known and expected phenomenon [29], as even the slightest deviation from the expected dataset becomes statistically significant when sample sizes are large.

Thus all $p$- values will become 0 unless the observed data *exactly* recaptures the expected data, and will no longer be useful as a measure of data quality.

Instead, it is possible to simply observe visually in the 100M sample case that, while the sampled distribution closely follows the expected distribution, there is a clear discontinuity between integer values of 127 and 128. This corresponds to an error in the most significant bit of the integers, which is the first bit determined using the probability-tree structure. Close inspection of the 100M curve reveals further, smaller discontinuities at other locations.

To understand the source of these discontinuities, the sensitivity of the sampled distribution as a result of variations at each individual level of the probability tree is investigated by using a single, ideal device for seven of the levels and a varied device for the eighth. This process is repeated for each level, with results on the sampled distributions corresponding to the variations in each of the three highest levels of the probability tree shown in Fig. 6a-c. Fig. 6a shows the largest positive and largest negative deviations in the device thermal stability for the most significant bit, affecting the biasing of coinflips at the top level of the probability-tree. It is seen that when the deviation is large and positive, integers between 0 and 127 are under-sampled while integers between 128 and 255 are over-sampled; the reverse is true when the deviation is negative. When the second bit is varied as in Fig. 6b, the sequence of alternating over- and under-estimates divides the domain equally into four segments. The pattern continues when varying the third bit in Fig. 6c, although with each successive level down the probability-tree the size of the deviations from the exponential distribution shrink. The discontinuities seen in the 100M dataset in Fig. 5 are thus the result of the summation of these variation patterns over all levels in the probability-tree; as the effect is most pronounced for variations at the top of the tree, the discontinuity is strongest at the center of the distribution. This implies that the quality of the first flip is the most critical to accurately sampling from distributions using the probability-tree.

## 6. Application

As an additional test of the quality of tRNG using an MTJ via the tuned probability-tree framework, the MTJ-sampled datasets are used during the numerical solution of physical phenomenon. Specifically, random walk solutions to partial integro-differential equations describing a simplified problem in particle physics transport are computed. The particular problem solved is a one-dimensional, initial-value, time-dependent problem describing the angular flux density $\Phi(t,\Omega)$ of a particle in an absorbing medium, with directional states $\Omega \pm 1$. The particle can either change states according to a Poisson process with some rate $\sigma_s$, or be absorbed according to another Poisson process with rate $\sigma_a$. While there are analytic expressions which describe this process, its solution can be cast in a probabilistic representation and the solution estimated in a Monte Carlo fashion by averaging the result of many random walks. For full details on the simulated equations and probabilistic representation, see note SN3.1 in the supplementary material to [30]. The solution requires both exponentially and uniformly distributed random variables, which can each be sampled using the MTJ-based device in a tuned probability-tree framework.

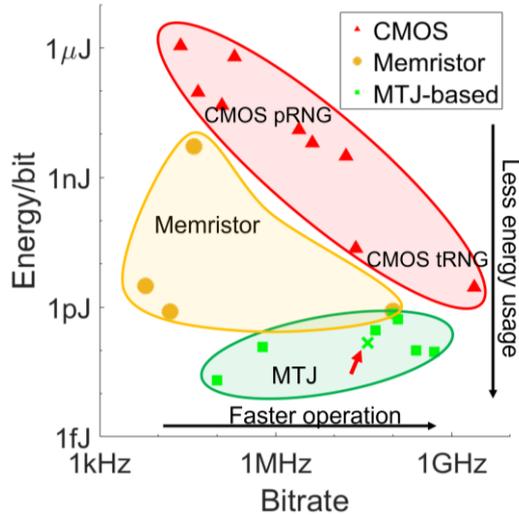

Fig. 8. Benchmarking random bit generation for CMOS, Memristor, and MTJ based technologies. Data is reported in [1] as time and energy per integer, and so converted to bitwise data for plotting purposes. This work is indicated by the red arrow; it does not include energy or delay effects needed to generate SOT pulse in external CMOS circuitry. See Table 2 for the underlying data.

TABLE II
RNG BENCHMARKING

| Ref. | Technology | Energy/bit [pJ] | Bitrate [Mb/s] |
|---|---|---|---|
| [1] | CMOS pRNG | $1.6E^4 - 1.1E^6$ | 0.024 -16 |
| [34] | CMOS tRNG | 2.9 | 2400 |
| [35] | CMOS tRNG | 23 | 231.6 |
| [32] | Memristor SNG | 0.81 | 100 |
| [36] | Memristor SNG | 5230 | 0.04 |
| [37] | Memristor SNG | 0.8 | 0.016 |
| [38] | Memristor SNG | 3.15 | 0.006 |
| [33] | MTJ SNG | 0.526 | 122 |
| [39] | SMART | 0.092 | 500 |
| [40] | SMART | 0.292 | 50 |
| [41] | Superparamagnetic | 0.02 | 0.1 |
| [42] | Spin Dice | 0.122 | 0.6 |
| [43] | Precessional | 0.1 | 244 |
| [44] | Parallel MTJs | 0.7 | 100 |
| This work | SOT-Pulsed | 0.15 | 37 |

For this demonstration, scattering rate $\sigma_s = 0.5$ and $\sigma_a = 5.0$. Thus, the sampled exponential distribution has parameter $\lambda = (\sigma_s + \sigma_a) = 5.5$. As 99.9983% of all expected sampled values fall in the interval $[0,2]$ for this exponential, the initial lower and upper bounds to truncate the distribution domain are chosen to be 0 and 2, respectively. In order to generate sampled numbers on the order of double-precision, this requires 55 consecutive flips for each sampled draw, as $\frac{2}{2^{55}} \approx 5.55E^{-17} < 1.0E^{-16}$. Only 54 consecutive flips are required for each uniformly sampled number, as the uniform distribution is supported on [0,1] (and thus its bounds are 0 and 1). In total, 128k random walks are performed; as each random walk solution requires ~80k exponential draws and ~150k uniform draws, this necessitates ~10 million exponential samples and ~20 million uniform samples for the entire simulation. The MTJs used to generate samples have normally-distributed variations in their thermal stability according to Section 4 and MTJ and simulation parameters follow those listed in Table I. The averaged results of the random walk simulations are compared to the analytical solution in Fig. 7a, showing excellent agreement.

To demonstrate the importance of careful construction of the probability-tree via proper truncation of the distribution domain the experiment is repeated to instead sample the exponential distribution on the interval [0, 1], which captures only 99.59% of the expected sampled values. This requires one fewer flip for each exponential sample to achieve 16-bit precision. The results of 128k random walks is shown in Fig. 7b, where it can be seen that both the $\Omega = +1$ and $\Omega = -1$ lines are now beneath the analytical solution by about 5%. The likely reason for this is the absence of rare, large draws push the flux value upwards during the simulation. The necessary size of the sampled distribution domain and precision of the sampled numbers will depend on the problem being solved, but can be controlled via the correct choice of initial lower and upper domain bounds and number of probability-tree levels.

In addition to the ability to generate large sets of random draws to accurately capture the physics of the particle transport problem, it is worth noting the time and energy usage savings to perform the computation as compared to traditional CMOS pRNGs. There are three stages to each individual MTJ coinflip: a pulse, a relaxation, and an MTJ read. With the given parameters in Table I each coinflip requires 25 ns total for pulsing and relaxation, and draws on average 0.15 pJ of energy (the energy draw does not depend significantly on the weighting [9]). Assuming a 2 ns read stage [31] using a 10 mV Ohmic read, the total time for a single coinflip is 27 ns; the energy of the read is significantly less than the pulse and relax, and so is ignored. This calculation does not include the energy to generate the pulse, which will come from external CMOS circuitry. In the particle transport example, each exponentially distributed random number requires 55 total coinflips, and thus takes 1.49 $\mu$s and 8.25 pJ to generate. In contrast, CMOS pRNG of 16-bit integers useful for cryptographic purposes can take anywhere from ~6-600 $\mu$s depending on platform and scheme, and draw no less than 0.35 [$\mu$J] per integer [1]. Further, in [2] 200,000 Gaussian-distributed random numbers are generated on a Nvidia GTX 970 GPU in ~20 [ms] using the Polar Box-Muller method, or 0.1 ns per number. While this is faster than the MTJ-based device in the probability tree, it is noted that the Nvidia GTX 970 parallelizes pRNG across up to 1664 cores, while the MTJ is a single device; operating many MTJs in parallel will allow for much faster tRNG as each random number can be generated by an independent MTJ, speeding up tRNG by a factor proportional to the number of devices. It is also possible to compare to other types of Stochastic Number Generation (SNG). [32] uses a two-terminal memristor device integrated with 65-nm CMOS technology to achieve 1 bit per 10 ns and 0.8 pJ per bit. Extrapolated to 55 generated bits this would be 55 ns and 44 pJ, slightly better than the SOT MTJ device presented here, though is unable to be dynamically tuned. And finally, [33] used HSPICE to model a MTJ for (voltage pulse) biased-SNG. It found it was able to do so at a rate of 1 bit per 8.21 ns with an energy of 0.53 pJ, with similar interpretation as the previous SNG example. For a visual comparison of the reported MTJ-based device energy usage and bitrate to CMOS pRNG, Memristor SNG, and a selection of other MTJ-based devices, see Fig. 8 and Table II. While the MTJ-based devices all produce random bits at much lower energy than either memristor SNG or CMOS technology, it is also useful to compare between the MTJ-based devices. Specifically, the superparamagnetic low-barrier magnet MTJ in [41] has the lowest energy usage, but is slower and susceptible to thermal and process variation due to the much lower thermal stability. [39] and [40] use medium-barrier magnets, which also have lower thermal energy barrier ($\sim$20-40$k_BT$) than this work. [42] uses high-barrier magnets ($109k_BT$), but with slower bitstream generation as a result. [43] generates bits more quickly than this work, but is difficult to tune as the stochastic behavior is due to precessional nature of the device. And [44] uses parallel MTJ-devices to improve bitstream quality, at higher energy cost. Thus, SOT MTJ in a tunable probability-tree is both faster and more energy efficient than modern CMOS technologies, and competitive with other non-CMOS stochastic devices with the additional ability to be dynamically tuned.

## 7. Conclusion

This paper studies how MTJs can function as tunable probability-trees for the purpose of generating random samples from arbitrary probability distribution functions. Through repeated numerical experiments and observing $p$-value distribution behavior, it was shown that the MTJ could reasonably draw integer samples from an exponential distribution with a given rate parameter. Through varying the MTJ thermal stability $\Delta$, it was shown that device parameters distort the ideal S-curve of an individual device and cause the device samples to become biased away from the desired distribution. However, by choosing to draw new samples from random devices and choosing the ideal S-curve such that there are normally distributed device variations, it is possible to draw numbers of the same quality as those drawn from an ideal device. It was also shown that the device variations induce different amounts of bias at different levels of the probability-tree, with the highest level of the tree being most sensitive to device variations.

The probability-tree framework was used to generate exponentially and uniformly distributed numbers for Monte Carlo-type simulation of a particle's angular flux density for a problem in particle physics transport. Generating 30M total random samples, the solution generated using the MTJ-based device showed excellent agreement with the analytical solution. Further, the random number generation was achieved at significantly lower energy usage than traditional CMOS pRNG, and at comparable energy usage and bitrate as other MTJ-based tRNG. The proper construction of the probability-tree, via correct identification of the distribution domain bounds and level of precision required, was demonstrated in this problem by over-truncating the distribution domain, causing the MTJ-based device solutions to deviate from the analytical solution.

## Acknowledgements

The authors acknowledge support from the DOE Office of Science (ASCR/BES) Microelectronics Co-Design project COINFLIPS. S.L. acknowledges support from the National Science Foundation graduate research fellowship program under Grant No. 2021311125. The authors acknowledge the Texas Advanced Computing Center (TACC) at the University of Texas at Austin, Austin, TX, USA, for providing the High Performance Computing resources that have contributed to the research results reported in this article. URL: http://www.tacc.utexas.edu. This article describes objective technical results and analysis. Any subjective views or opinions that might be expressed in the article do not necessarily represent the views of the U.S. Department of Energy or the United States Government. Sandia National Laboratories is a multimission laboratory managed and operated by the National Technology and Engineering Solutions of Sandia, LLC, a wholly owned subsidiary of Honeywell International Inc., for the U.S. Department of Energy's National Nuclear Security Administration under Contract DE-NA0003525.